# Fluorine solubility and superconducting properties of Sm(O,F)BiS$_2$ single crystals

*Koki Kinami[1], Yuji Hanada[1], Masanori Nagao[1,*], Akira Miura[2], Shigeto Hirai[3], Yuki Maruyama[1], Satoshi Watauchi[1], Isao Tanaka[1]*

[1]*University of Yamanashi, 7-32 Miyamae, Kofu, Yamanashi 400-0021, Japan*

[2]*Hokkaido University, Kita-13 Nishi-8, Kita-ku, Sapporo, Hokkaido 060-8628, Japan*

[3]*School of Earth, Energy and Environmental Engineering, Kitami Institute of Technology, 165 Koen-cho, Kitami, Hokkaido 090-8507, Japan*

[*]Corresponding Author

Masanori Nagao

Postal address: University of Yamanashi, Center for Crystal Science and Technology Miyamae 7-32, Kofu, Yamanashi 400-0021, Japan

Telephone number: (+81)55-220-8610

Fax number: (+81)55-254-3035

E-mail address: mnagao@yamanashi.ac.jp

**Abstract**

Sm(O,F)BiS$_2$ superconducting single crystals with various nominal F contents have been grown using KI-KCl flux. The solid solution limit of F at the O-site in Sm(O,F)BiS$_2$ single crystals was approximately less than 20 at%. F concentrations of Sm(O,F)BiS$_2$ single crystals were hardly controlled by nominal F contents. It suggests the existence of intrinsic stable phase. Superconductivity was shown at around 5 K. Superconducting transition temperature with zero resistivity were increased with increasing nominal F contents within less than 70 at% in O-site, despite similar F concentration and *c*-axis lattice constants.

**Main text**

**1. Introduction**

Superconductors with high transition temperature ($T_c$) are composed of an alternate stacking of a superconducting layer and a blocking layer structure, such as cuprate superconductors [1-3], and iron-based superconductors [4,5]. $R$(O,F)BiS$_2$ ($R$: rare earth elements) superconductors also show similar stacking structure [6]. However, superconducting transition exhibits low temperature which is 2-5 K range [7]. Superconducting transition temperatures ($T_c$) of $R$(O,F)BiS$_2$ are enhanced by chemical pressure, which is induced by substituting $R$-site elements with lanthanoid elements of smaller ion radii [7,8]. We focused on Sm substitution in $R$-site for $R$(O,F)BiS$_2$, which ionic radius is smaller than that of other lanthanoid elements, such as La, Ce, Pr and Nd [6,9-11]. In consequence, Sm(O,F)BiS$_2$ superconducting single crystals with approximately 17 at% F substituted in O-site were successfully grown using KI-KCl flux [12]. By G. S. Thakur *at al.*, Sm(O,F)BiS$_2$ single crystals were successfully grown using CsCl-KCl flux [13]. In our experiment, exceedingly tiny Sm(O,F)BiS2 single crystals with the size of less than 20 μm and impurity phases such as Bi$_2$S$_3$ were obtained using CsCl-KCl flux. Perhaps, this reason is due to the molar ratio of CsCl-KCl flux. The CsCl : KCl molar ratio of our experiment and G. S. Thakur *at al.* experiment were 5 : 3 and 1 : 1, respectively. Our previous paper showed that Sm(O,F)BiS$_2$ single crystals had been grown using KI-KCl flux [12]. Therefore, we employed KI-KCl flux for Sm(O,F)BiS$_2$ single crystals growth.

In this paper, we are interested in the relationship between F contents and superconducting properties for Sm(O,F)BiS$_2$ single crystals. We report the growth of Sm(O,F)BiS$_2$ single crystals with various nominal F contents. Analytical compositions, *c*-axis lattice constants and superconducting properties of the obtained Sm(O,F)BiS$_2$ single crystals with various nominal F contents are investigated.

## 2. Material and methods

Sm(O,F)BiS$_2$ single crystals were grown by using KI-KCl flux. The raw materials: Sm$_2$S$_3$ (99.9 % : Mitsuwa Chemicals Co., Ltd.), Bi$_2$S$_3$ (99.99 % : Kojundo Chemical Lab. Co., Ltd.), Bi$_2$O$_3$ (99.9 % : Kojundo Chemical Lab. Co., Ltd.), BiF$_3$ (99.99 % : Rare Metallic Co., Ltd.), and Bi (99.999 % : Rare Metallic Co., Ltd.) were weighted to achieve the nominal composition of $SmO_{1-x}F_xBiS_2$ ($x$ = 0.2-0.8). The molar ratio of the KI-KCl flux was KI:KCl = 3:2 using KI (99.5 % : Kanto Chemical Co., Inc.) and KCl (99.5 % : Kanto Chemical Co., Inc.). The mixture of raw materials (0.8 g) and KI-KCl flux (5.0 g) were ground using a mortar and pestle, and then sealed into an evacuated quartz tube (~10 Pa). The quartz tube was heated at 700 °C for 10 h, and subsequently cooled to 600 °C at the rate of 0.5 °C/h. Then, the sample was cooled to room temperature, while still in the furnace. The heated quartz tube was then opened in air, and the obtained materials were washed and filtered using distilled water to remove the KI-KCl flux. Single crystals growth condition except to the F contents is same to previous reports [12].

X-ray diffraction (XRD, MultiFlex, Rigaku) using Cu K$\alpha$ radiation was employed to identify the crystal structure and determine their orientation. The lattice constants were determined using X-ray diffraction with a two dimensional detector (Bruker D8). The compositional ratio of the obtained Sm(O,F)BiS$_2$ single crystals was evaluated using energy dispersive X-ray spectrometry (EDS, Quantax 70, Bruker), and the microstructure was observed using scanning electron microscopy (SEM) (Hitachi

High-Technologies, TM3030). The obtained compositional values were normalized using the atomic content of S =2.00, while Sm and Bi were measured. On the other hand, the atomic ratio of O and F was determined using electron probe microanalysis (EPMA, JXA-8200, JEOL). Those standard samples were $Bi_2O_3$ and $PbF_2$ for O and F, respectively. The number of analyzed points was around three to five in the crystal. The estimated values were normalized using the atomic content of O + F = 1.00. Atomic content of each element is defined as $C_{XX}$ (XX: symbol of element). In particular, the normalized analytical F content is defined as "$y$" which was same meaning of $C_F$. Actually, F contents were dramatically different between nominal and analytical composition, which were defined as "$x$" and "$y$", respectively.

The valence states of the Sm components in the obtained single crystals were estimated by X-ray absorption spectroscopy (XAS) analysis of Sm-$L_3$ edges using an Aichi XAS beam-line with a synchrotron X-ray radiation (BL05S1: Experimental No.201905108).

The resistivity-temperature ($\rho$-$T$) characteristics of the obtained single crystals were measured using the standard four-probe method, in constant current density ($J$) mode, employing the physical property measurement system (PPMS DynaCool, Quantum Design). The electrical terminals were fabricated using Ag paste. The superconductivity-onset temperature ($T_c^{onset}$) is defined as the temperature where the normal conducting state deviates from linear behavior in the $\rho$-$T$ characteristics. The zero-resistivity critical temperature ($T_c^{zero}$) is defined as the temperature at which resistivity is lower than 20 $\mu\Omega$cm. The $T_c^{onset}$ with applied various magnetic fields were measured from the $\rho$-$T$ characteristics under a magnetic fields ($H$) perpendicular to the $c$-axis and parallel to the $c$-axis with range of 0.1-9.0 T, and then the upper critical field values of each magnetic fields direction, $H_{C2}^{\perp c\text{-axis}}$ and $H_{C2}^{//c\text{-axis}}$ were estimated from the linear extrapolations of those $T_c^{onset}$ data. The superconducting anisotropy ($\gamma_s$) was calculated from two kinds of methods which were ratio of upper critical field ($H_{C2}^{\perp c\text{-axis}}$ and $H_{C2}^{//c\text{-axis}}$) and effective mass model [14-16].

## 3. Results and discussion

The single crystals of Sm(O,F)BiS$_2$ were grown, only in the range of $x$ = 0.3-0.8 of SmO$_{1-x}$F$_x$BiS$_2$. We tried to grow the $x$ = 0.2 sample. However, the Sm(O,F)BiS$_2$ single crystals were not obtained, whereas $Bi_2S_3$ crystals and Sm-O-S-F impurities (Note: The typical atomic composition of Sm-O-S-F impurity was detected to be roughly Sm:O:S = 2:1:1 with included small amount fluorine by EDS analysis. However, this impurity phase could not be identified.) were synthesized. Figure 1 shows a typical SEM image of $Sm(O,F)BiS_2$ single crystals. The obtained $Sm(O,F)BiS_2$ single crystals have a platy-shape with size of 0.3-0.5 mm and thickness of 10-20 μm, and side faces are ill-defined. According to XRD data, $c$-plane is well-developed. In other words, the grown crystals have the $c$-axis perpendicular to the $c$-plane. The $a$-axis lattice constants were also tried to measure using X-ray diffraction with a two dimensional detector. However, they were not able to be estimated due to their particularly thin character along the $a$-axis direction in the obtained single crystals. X-ray powder diffraction was also tried using the crushed $Sm(O,F)BiS_2$ single crystals. However, reactions yield of $Sm(O,F)BiS_2$ single crystals was a few mass% which was too little amount for X-ray powder diffraction measurement. Main phase, analytical compositions and $c$-axis lattice constants of the obtained single crystals are listed in Table I. The atomic ratios of Sm, Bi and S approximately corresponded with nominal compositions in all samples, respectively. K, I and Cl from the flux were not detected in the single crystals at the minimum sensitivity limit of 0.1 wt%. On the other hand, averaged values of analytical F contents ($y$) of $x = 0.3$ and $x = 0.4$-$0.8$ were $y = 0.13$ and $y = 0.16$-$0.19$, respectively (Table I). This result suggests that the solid solution limit of F at the O site is approximately less than 20 at% in the Sm(O,F)BiS$_2$ single crystals. On the other hand, the lower limit of F may be approximately 13 at%. However, those values have a large margin of error using EPMA, and then analytical F contents of the all samples become same within the error range. Moreover, differences of the $c$-axis lattice constants in all samples were not observed within the error range. These results suggest that approximately 20 at% F substituted Sm(O,F)BiS$_2$ has possibility to be of intrinsic stable phase. Therefore, F concentrations and $c$-axis lattice constants of Sm(O,F)BiS$_2$ single crystals are hardly controlled by nominal F contents.

Then, the valence states of the Sm components in the obtained single crystals were evaluated by XAS. Figure 2 shows the Sm $L_3$-edge absorption spectra of the Sm(O,F)BiS$_2$ single crystals with $x = 0.3$ and 0.5 at room temperature obtained by XAS analysis. Sm $L_3$-edge showed a peak at 6719 eV, which can be assigned to a trivalent electronic configuration ($Sm^{3+}$). This is consistent with the other XAS results for $Sm^{3+}$ [17]. The associated absorption-edge spectra of both samples ($x$ = 0.3 and 0.5) show similar peak shape. This result suggests the same valence state that is characteristic for mainly trivalent electronic configuration ($Sm^{3+}$).

The temperature dependence of the electrical resistivity ($\rho$-$T$) for obtained Sm(O,F)BiS$_2$ single crystals were shown in Figure 3. All samples exhibited metallic behavior, and superconductivity was shown at around 4.5 K. Superconducting transition temperatures with zero resistivity ($T_c^{zero}$) are increased with increasing nominal F contents within less than $x$ = 0.7, and then shift to decrease within more than $x$ = 0.8. The superconducting transition temperatures within $x$ = 0.3-0.8 range changed despite similar analytical F contents and $c$-axis lattice constants. If superconducting transition temperatures of those samples were the same, they should decrease with increasing the measurement current densities ($J$). However, those superconducting transition temperatures have no clear relation with the experimental current densities ($J$) which were shown in Table II. Therefore, the change of superconducting transition temperatures may be originated from other properties of the obtained single crystals.

For comparison, the superconducting transition width with $x$ = 0.5, 0.7 and 0.8 are broader than those of $x$ = 0.3, 0.4 and 0.6. The main reason for this observation could be the inhomogeneous F distribution in the single crystals. More specifically, these results indicate that superconducting transition temperatures are depended on the nominal F contents. At this moment, the origin of various superconducting transition temperature is unclear. Although, we predict that $a$-axis lattice constants or crystallinity are dominant in superconducting transition temperature for $Sm(O,F)BiS_2$ single crystals.

The $\rho$-$T$ characteristics of the obtained $Sm(O,F)BiS_2$ single crystals with $x$ = 0.3 below 10 K under magnetic fields ($H$) of 0.1-9.0 T, which were perpendicular to the $c$-axis ($H \perp c$-axis) and parallel to the $c$-axis ($H // c$-axis) are presented in Figs. 4(a) and (b), respectively. The suppression of superconducting transition temperature under the effect of $H // c$-axis was more significant than that attributed to the $H \perp c$-axis. $T_c^{onset}$ was estimated from Fig. 4, and the field dependences of $T_c^{onset}$ under the effects of the $H \perp c$-axis and $H // c$-axis are plotted in Fig. 5. From the linear extrapolations of the $T_c^{onset}$ data, the upper critical field values, $H^{\perp c\text{-axis}}_{C2}$ and $H^{// c\text{-axis}}_{C2}$, were predicted to be 35 and 1.1 T, respectively. The upper critical field values of another nominal F contents samples were also estimated using same way. The upper critical field values of each magnetic fields direction, $H^{\perp c\text{-axis}}_{C2}$ and $H^{//c\text{-axis}}_{C2}$ for all samples were estimated to be 29-43 T and 1.0-1.4 T, respectively. These results indicate that the obtained Sm(O,F)BiS$_2$ single crystals exhibit high superconducting anisotropy. We determined the superconducting anisotropy ($\gamma_s$) using the ratio of upper critical field ($H^{\perp c\text{-axis}}_{C2}$ divided by $H^{//c\text{-axis}}_{C2}$) or an effective mass model [14]. $\gamma_s$ was calculated by the upper critical fields according to Eq. (1):

$$\gamma_s = H^{\perp c\text{-axis}}_{C2}/H^{//c\text{-axis}}_{C2}. \qquad (1)$$

On the other hand, $\gamma_s$ was also evaluated using the effective mass model [14]. The angular ($\theta$) dependence of resistivity ($\rho$) was measured at various magnetic fields ($H$) values in the vertices liquid state to estimate $\gamma_s$ of the obtained single crystals [15,16]. The reduced field ($H_{red}$) was evaluated using Eq. (2) for the effective mass model:

$$H_{red} = H(\sin^2\theta + \gamma_s^{-2}\cos^2\theta)^{1/2}, \qquad (2)$$

where $\theta$ is the angle between the $c$-plane and magnetic field [14]. The $\gamma_s$ value was estimated using the best scaling of the graph illustrating the $\rho$-$H_{red}$ relationship. Figure 6(a) shows the angular ($\theta$) dependence of resistivity ($\rho$) at various magnetic fields ($H =$

0.1-9.0 T) in vertices liquid state (3 K) for the grown $Sm(O,F)BiS_2$ single crystal with $x$ = 0.3. The $\rho$-$\theta$ curve exhibited almost two-fold symmetry. The $\rho$-$H_{red}$ scaling obtained from Fig. 6(a) using Eq. (2) is presented in Fig. 6(b). The best scaling was obtained for $\gamma_s$ = 20, as shown in Fig. 6(b). However, this plot was deviated from the scaling at higher magnetic fields. The $\gamma_s$ from effective mass model of another nominal F contents samples were also evaluated using the same way, and then they exhibited similar behavior. For the obtained $Sm(O,F)BiS_2$ single crystals, $\gamma_s$ values from the ratio of upper critical fields were estimated as higher than those obtained from effective mass model. These values were shown in Table II. The $\gamma_s$ values showed the same trend as in the previous report [12]. Measurement current densities ($J$), superconducting transition temperature ($T_c^{onset}$ and $T_c^{zero}$), upper critical fields ($H_{C2}^{\perp c\text{-axis}}$ and $H_{C2}^{//c\text{-axis}}$) and superconducting anisotropies $\gamma_s$ (determined from upper critical field and effective mass model) are listed in Table II. In a conventional (BCS-like) superconductor at the weak-coupling limit, the Pauli limit ($H_p$) is calculated to be 9.2-10.1 T, since $H_p$ = 1.84 $T_c$, ($T_c^{onset}$ = 5.0-5.5 K) [18]. Thus, the upper critical field in the $c$-plane ($H_{C2}^{\perp c\text{-axis}}$ = 29-46 T) is significantly higher than the Pauli limit ($H_p$ = 9.2-10.1 T), indicating the possibility of an unconventional superconductor. On the other hand, the upper critical fields and superconducting anisotropies have no observed correlations within $x =$ 0.3-0.8 range, but the values of those superconducting parameters exhibited some differences. The obtained chemical state such as $c$-axis lattice constants and analytical F contents were similar within those nominal F contents ranges. The origin of the differences of those superconducting parameters among each nominal sample F contents may be other chemical parameters. At a minimum, analytical F contents had similar value which was approximately 20 at%. It strongly suggests the possibility of the existence of intrinsic stable phase for $Sm(O,F)BiS_2$. Thus this predicts that approximately 20 at% F substituted structure is fairly stable phase in $Sm(O,F)BiS_2$ single crystals. However, the proposition of the relationship between superconducting properties and chemical states for $Sm(O,F)BiS_2$ single crystals was also given. Further characterizations of $Sm(O,F)BiS_2$ single crystals are therefore necessary to clarify this proposition. Additional investigations such as $a$-axis lattice constants and elemental deficiency are also required.

## 4. Conclusions

We have successfully grown Sm(O,F)BiS$_2$ single crystals with the nominal compositions of SmO$_{1-x}$F$_x$BiS$_2$ ($x$ = 0.3-0.8) using KI-KCl flux. The solid solution limit of F at the O site is approximately less than 20 at%. The difference of $c$-axis lattice constants could not be observed in the all samples. But analytical F contents were similar value which was approximately 20 at% F substitution in O-site. Therefore, Sm(O,F)BiS$_2$ with approximately 20 at% F substitution has possibility to be of intrinsic stable phase. In consequence, F concentrations and $c$-axis lattice constants are hardly controlled by nominal F contents. On the other hand, the peak shape of associated absorption-edge spectra for Sm was almost the same between $x$ = 0.3 and 0.5, which was mainly trivalent electronic configuration (Sm$^{3+}$). Superconducting transition temperature were increased with increasing nominal F contents within less than $x$ = 0.7 despite similar chemical properties. The upper critical fields and superconducting anisotropies showed no remarkable tendencies regardless of the various nominal F contents within $x$ = 0.3-0.8 range. However, values of those superconducting parameters exhibited some differences. Those observations may be originated from other chemical parameters such as *a*-axis lattice constants, elemental deficiency and so on.

**Acknowledgments**

 The XAS experiments were conducted at the BL05S1 of Aichi Synchrotron Radiation Center, Aichi Science & Technology Foundation, Aichi, Japan (Experimental No.201905108). This research was partially supported by Grants-in-Aid for Scientific Research (C) (JSPS KAKENHI Grant Number JP19K05248).

**Figure captions**

Figure 1. Typical SEM image of Sm(O,F)BiS$_2$ single crystal.

Figure 2 (Color online). XAS spectra at room temperature for the obtained Sm(O,F)BiS$_2$ single crystals with $x$ = 0.3, 0.5 and standard valence samples at the Sm $L_3$-edge.

Figure 3 (Color online). (a) Temperature dependencies of the electrical resistivity for obtained Sm(O,F)BiS$_2$ single crystals. (b) Enlargement of superconducting transition in the temperature range of 3-6 K, normalized with resistivity at 6 K.

Figure 4. Temperature dependence of resistivity for the Sm(O,F)BiS$_2$ single crystal with $x$ = 0.3 under the magnetic fields ($H$) of 0.1-9.0 T (a) perpendicular to the $c$-axis ($H \perp$ $c$-axis) and (b) parallel to the $c$-axis ($H // c$-axis).

Figure 5. Date in figure 4 after plotting of field dependences of $T_c^{onset}$ under the magnetic fields ($H$) perpendicular to the $c$-axis ($H \perp c$-axis) and parallel to the $c$-axis ($H$ // $c$-axis) The lines are linear fits to the date.

Figure 6. (a) Angular $\theta$ dependence of resistivity $\rho$ for the Sm(O,F)BiS$_2$ single crystal with $x = 0.3$ in vertices liquid state (3 K) at various magnetic fields (0.1-9.0 T). (b) Date in figure 6 (a) after best scaling of angular $\theta$ dependence of resistivity $\rho$ at a reduced magnetic field of $H_{red} = H(\sin^2\theta + \gamma_s^{-2}\cos^2\theta)^{1/2}$.

Table I. Main phase, analytical compositions and $c$-axis lattice constants of the obtained single crystals grown from the nominal compositions of $SmO_{1-x}F_xBiS_2$.

| $x$ | 0.2 | 0.3 | 0.4 | 0.5 | 0.6 | 0.7 | 0.8 |
|---|---|---|---|---|---|---|---|
| Phase | $Bi_2S_3$ | \multicolumn{6}{c}{$Sm(O,F)BiS_2$} | | | | | |
| $C_F$ ( $= y$) | --- | $0.13_{\pm 0.02}$ | $0.18_{\pm 0.00}$ | $0.16_{\pm 0.04}$ | $0.19_{\pm 0.07}$ | $0.19_{\pm 0.06}$ | $0.19_{\pm 0.05}$ |
| $C_{Sm}$ | --- | $1.06_{\pm 0.11}$ | $1.00_{\pm 0.09}$ | $1.05_{\pm 0.04}$ | $1.04_{\pm 0.03}$ | $0.99_{\pm 0.12}$ | $1.02_{\pm 0.04}$ |
| $C_{Bi}$ | --- | $0.99_{\pm 0.02}$ | $0.97_{\pm 0.04}$ | $0.96_{\pm 0.03}$ | $0.95_{\pm 0.02}$ | $0.98_{\pm 0.04}$ | $0.95_{\pm 0.04}$ |
| $C_S$ | --- | \multicolumn{6}{c}{2.00} | | | | | |
| $c$-axis (Å) | --- | $13.43_{\pm 0.03}$ | $13.42_{\pm 0.02}$ | $13.42_{\pm 0.03}$ | $13.42_{\pm 0.02}$ | $13.42_{\pm 0.02}$ | $13.42_{\pm 0.02}$ |

Table II. Experimental current densities ($J$) and superconducting properties ($T_c$, $H_{C2}$, and $\gamma_s$) of the obtained single crystals grown from the nominal compositions of SmO$_{1-x}$F$_x$BiS$_2$.

| F | $x$ | 0.3 | 0.4 | 0.5 | 0.6 | 0.7 | 0.8 |
|---|---|---|---|---|---|---|---|
| | $y$ | 0.13$_{\pm 0.02}$ | 0.18$_{\pm 0.00}$ | 0.16$_{\pm 0.04}$ | 0.19$_{\pm 0.07}$ | 0.19$_{\pm 0.06}$ | 0.19$_{\pm 0.05}$ |
| $J$ (A/cm$^2$) | | 5.66 | 2.83 | 5.21 | 1.36 | 2.68 | 2.71 |
| $T_c$ (K) | onset | 5.0 | 5.2 | 5.5 | 5.5 | 5.5 | 5.3 |
| | zero | 4.2 | 4.2 | 4.4 | 4.7 | 4.7 | 4.4 |
| $H_{c2}$ (T) | $H \perp c$-axis | 35 | 46 | 29 | 43 | 34 | 37 |
| | $H // c$-axis | 1.1 | 1.4 | 1.1 | 1.1 | 1.3 | 1.0 |
| $\gamma_s$ | ratio of $H_{c2}$ | 32 | 33 | 26 | 39 | 26 | 37 |
| | effective mass | 20 | 20 | 10 | 35 | 20 | 15 |

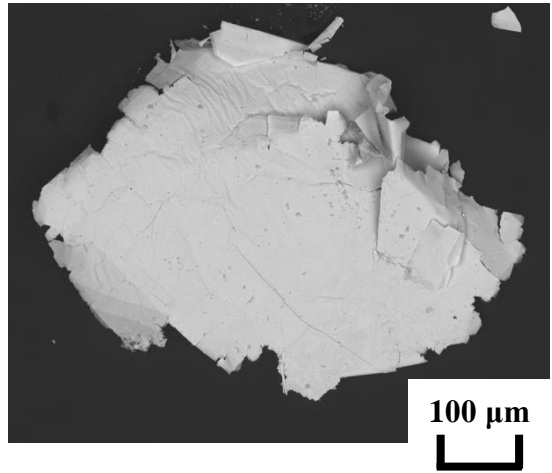

**Figure 1.**

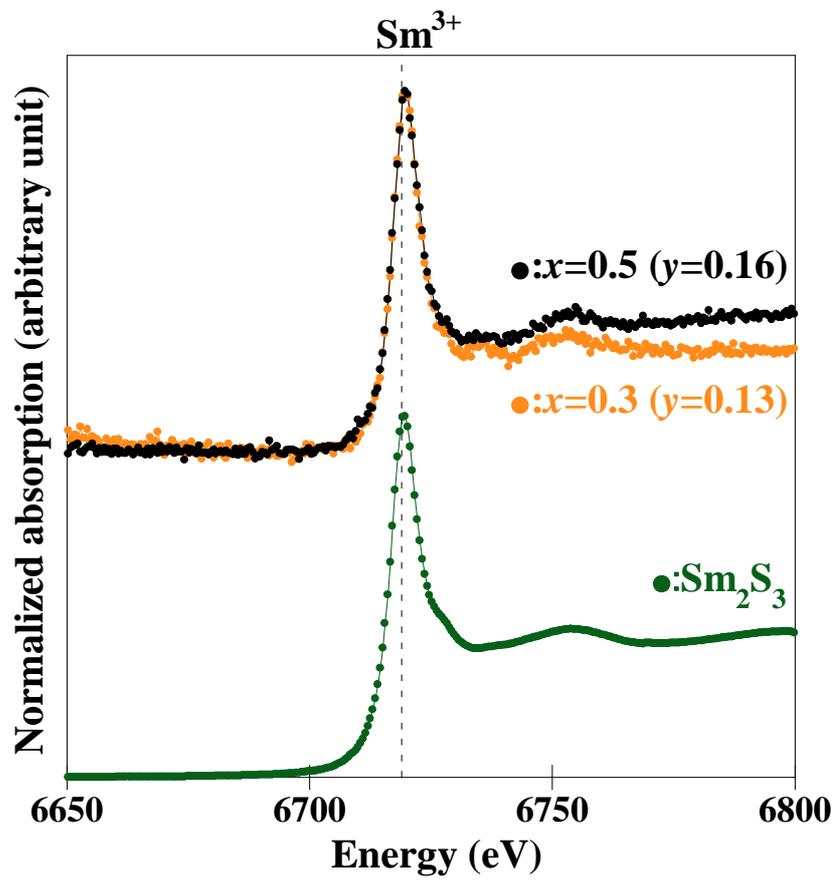

**Figure 2 (Color online).**

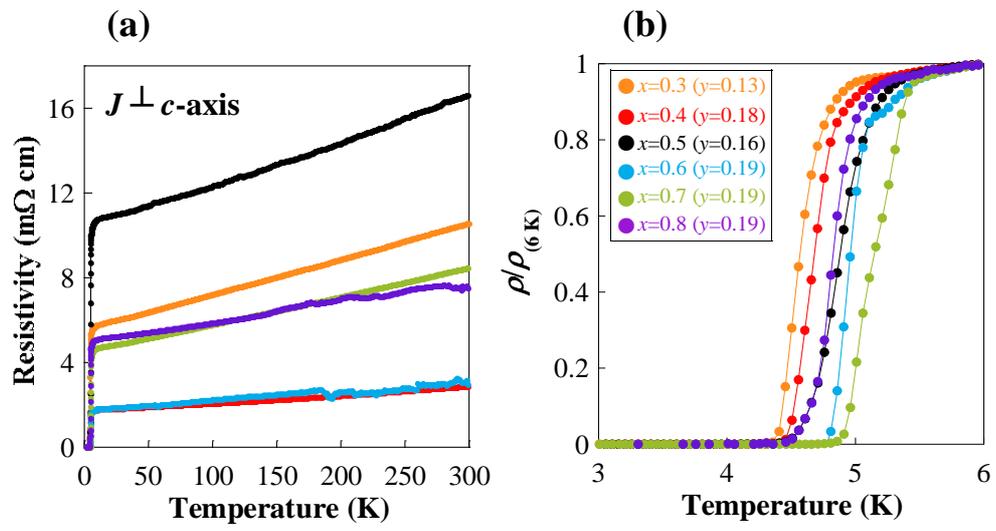

**Figure 3 (Color online).**

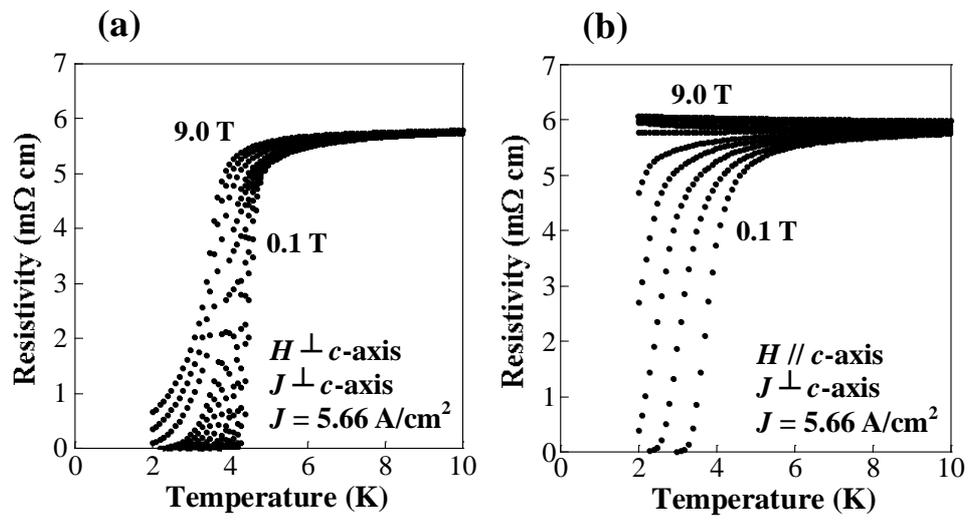

**Figure 4.**

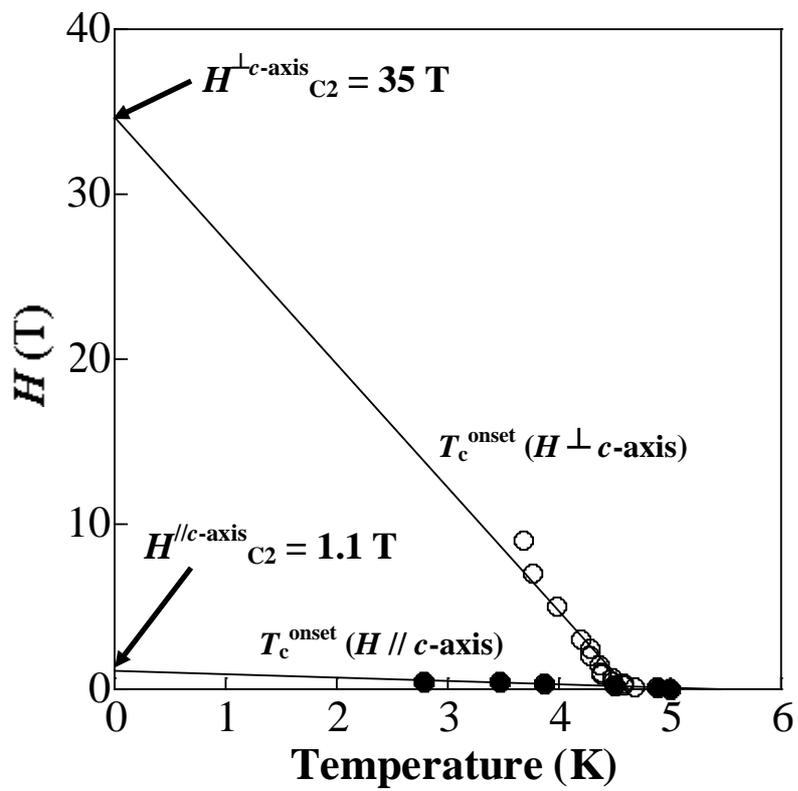

**Figure 5.**

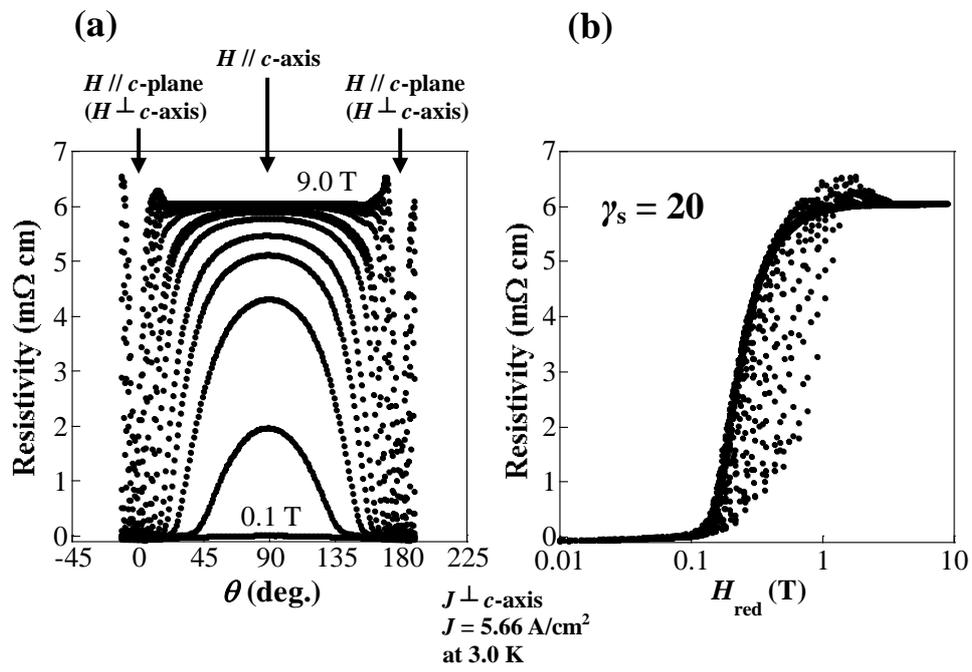

**Figure 6.**